\documentstyle[12pt,epsf]{article}
\def\beq{\begin{equation}}
\def\eeq{\end{equation}}
\begin{document}
\begin{center}
{\Large \bf
  Germanium Detector with Internal Amplification for Investigation of
Rare Processes } \\
\vspace{2.0cm}
{\large A.S Starostin, A.G.Beda }
\end{center}

\begin{center}
{\normalsize  {SSC~ RF~ ITEP,~ Moscow~ 117259,~ Russia}}\\
\end{center}
\begin{abstract}
 Device of new type is suggested - germanium detector with internal
amplification. Such detector having effective threshold about 10 eV
 opens up fresh opportunity for investigation of dark matter, measurement
of neutrino magnet moment, of neutrino coherent scattering at
nuclei and for study of solar neutrino problem.
 Construction of germanium detector with internal amplification
and perspectives of its use are described.
\end{abstract}

\section{Introduction}

 The detectors having low background and low threshold are required for
 investigation of rare processes involving low energy neutrino and weak
 interacting particles. Such detectors can be used effectively for search for
 the dark matter, for measurement of neutrino magnetic moment and
 coherent neutrino scattering by nuclei and for investigation of solar
 neutrino problem.

For above investigations one needs low background detector  of mass  several
$ kg$ and with threshold less than 1000 $eV$. The cryogenic and germanium
detectors partly correspond to these requirements. The drawback of the
 cryogenic detectors is the complexity of their production and use. The
drawback of germanium detectors is a rather high threshold $2\div10$ $KeV$ which
is due to a leakage current and electronic and microphonic noises. It would
be very attractive to provide effective decreasing of the detector threshold
 using the internal proportional amplification of signal.
Internal proportional amplification in the semiconductor detectors is
realized now in the silicon avalanche photodiodes (APD)
  ${\cite{pans,farr}}$~, where a gain of about $10^{2} \div 10^{4}$ is
achieved by avalanche multiplication of electrons at electric field $
(5\div6)\times 10^{5}$ $V/cm$ in narrow $ p-n$ junction with sensitive volume
several $ mm^{3}$. Below we shall demonstrate the possibility of realizing
the germanium detector with internal amplification (GDA) and present its
design.

\section{Principles of GDA and its design.}

 In semiconductor detectors ( as in a gas proportional counter-PC, or
multiwire proportional chamber-MWPC) the conditions for internal
proportional amplification of electrons can be fulfilled.
It  is known that in APD critical electric field $E_{cr}$, that
provides multiplication of electrons at room temperature, is equal $
(5\div 6)\times10^{5}$ $V/cm$ . The $E_{cr}$ for germanium at
liquid nitrogen temperature can be defined from the dependence  of
electron drift velocity on electric field and energy of production of
electron-hole pairs and photons
${\cite{sakai}}$. For germanium at 77K
the magnitude of $E_{cr}$ derived in
this fashion is equal $9\times10^{4}$ $V/cm$. In APD and gas PC the critical
electric field is produced by a different way. In the first case $E_{cr}$
is achieved by high concentration of impurities in narrow junction. As a
  result the sensitive volume of APD is several ${mm}^3$. In gas PC the
$E_{cr}$ can be achieved by special configuration of the electric
field, due to large difference of cathode and anode sizes. In high purity
germanium (HPGe) with a sensitive volume near 100 $cm^3$
the $E_{cr}$ can be obtained by the same manner.

Electric field in the gas cylindrical PC is of the form:  \beq E(r)=
\frac{V}{r\ln \left( R_{2}/R_{1} \right) } \eeq where $V$ is applied voltage,
$R_{1}$ and $R_{2}$ are the radii of cathode and anode correspondingly, $r$
is a distance from anode.

One can see from
(1) that at $V=10^{3}$ $V$, $R_{1}=0.001$ $cm$ and $R_{2} = 1$ $cm$ $ E(r)$ is
about $10^{5}$ $V/cm$ near the anode.  Unlike gas PC the electric field in
the coaxial detector from high purity germanium is defined not only by
$V$, $R_{1}$ and $R_{2}$ but the concentration of donor (n-type) or acceptor
(p-type) impurities as well. The magnitude of the volume charge
in the sensitive volume of crystal depends on these impurities.  The electric
field in coaxial detector from HPGe with regards to the impurities
will be ${\cite{marl}}$ :

    \beq E(r)= \frac{N \,e}{2 \epsilon}\,r - \frac { \left [ V + ({ N \,e}/{ 4
    \, \epsilon}) \, \left ( R_2^2 - R_1^2 \right ) \right]} { r \ln{\left(
    R_2 / R_1\right) }}~~~, \eeq
where $N$ is impurity concentration, $e$ is the electron charge,$\epsilon$ is the
dielectric constant of germanium.

The electric field (2) can be expressed with a
 depletion voltage $V_{d}$, which is a minimum applied voltage
necessary to neutralize the volume charge and to provide the sensitive region
in the whole crystal volume.  The $ V_{d}$ for the coaxial detector,
taking into account that $R_{2} \gg R_{1}$, may be given by :
 \beq  V_{d} \approx - \frac {N \,q}{4 \epsilon} \, R_2^2 ~~~. \eeq
and the equation (2) may be rewritten for n-type germanium
in the final form :
 \beq  E(r)= -\frac {2 \,V_d} {R_2^2} \,r - \frac { V - V_d} { r \ln{
 \left( R_2 /R_1 \right) }}~~~.
 \eeq
 The dependence $E(r)$ on $r$ is shown on fig.1 for coaxial detector from
 HPGe.  Electric field near the anode reaches  $E_{cr}$ - magnitude required
 for avalanche multiplication of electrons. The coaxial germanium detector
 with internal amplification is the more appropriate for the low background
 spectrometers but the possibility of fabrication of inner electrode of 20
 micron radius is highly conjectural presently. So we shall consider the more
 realistic problem - fabrication of planar germanium detector with internal
 amplification, multistrip planar germanium detector, similar in design to
 MWPC.  The electric field in MWPC is of the form for one dimension case
 (the coordinates x and y relate to an centered on the wire, x is parrelel
 to the wire plane, y is perpendicular) :
      \beq
     E(0,y)=\frac{\pi \, V}{ s \left [ \frac{\pi \, L}{ s} \, - \ln{\pi\, d
     \over s} \right]} \coth{\pi \, y\over s} \eeq
 where $V$  is applied voltage, $s$ is  wire spacing , $d$ is a diameter of
 the wire and $L$
is the thickness of the planar detector. As in the case of coaxial
 germanium detector one must take into account for multistrip germanium
 detector  the depletion voltage $V_{d}$. In the case of planar
 germanium detector $V_{d}= -\frac{Ne}{2 \epsilon} \,L^2$.
 The electric field for multistrip germanium detector is of the form:
 \beq E(0,y) = -\frac{
   2 \,V_{d}}{L^2}\,y - \frac{ \pi \left( V -V_d \right )} {s \left [\frac
   {\pi \, L}{ s} \, - \ln{\pi\, d \over s} \right]} \coth{\pi \, y\over s}
     \eeq
where $d$ is the strip width. The dependence of $E(0,y)$ on $y$ for multistrip
germanium detector is
 shown on fig.2.  In the cases being considered the electric field near the
 anode is sufficient for avalanche multiplication of electrons ($E > 10^5$
 $V/cm$).  The amplification factor can be estimated as  $K=2^{h/l}$  where
 $l$ is a free electron path for inelastic scattering and $h$ is a length of
 avalanche region where $E>E_{cr}$.  The $l$ in germanium at
 77$K$ is equal 0.5 micron and for $L=3$ $cm$  $h$ is equal 5 micron
(see fig.2)
 so for this case it is possible to achieve $K=10^3$. If one does not need
high amplification factor it is possible to decrease $V$ or to increase the
 strip width $s$.

The GDA is assumed to use for investigation of rare processes so
 the spectrometer including GDA must have large mass of detector, it can be
 fabricated from separate modules of mass about 0.7 $kg$ each. One module
represents the multistrip planar germanium detector from HPGe  having
     impurity concentration less than $10^{10}$ $cm^{-3}$, measures
     $70\times70\times30$ $mm^{3}$ (see fig.3).  The 12 anode strips of 20
    micron width and of 65 $mm$ length are fabricated by photomask method
     ${\cite{gut}}$.  The cathode area is equal $65\times65$ $mm^2$ and the
fiducial volume is equal to 130 $cm^3$.  There are the guard electrodes in the
anode and cathode planes .  The anode strips can be connected together
     however it is more convenient to take signals from separate strips to
     suppress the background Compton gamma-quanta.

For the fabrication of GDA it is necessary to use the germanium crystals of
     uniform distribution of impurities to provide homogenous electric field
     near the anode. Second in importance it is the providing small depth and
 width of
     junction layer under the strips so the electric field near
     the strips is defined by junction dimensions.
     The design of GDA must gurantee the reliable cooling  of crystal since
the critical electric field and amplification factor  depend on free path of
  charge carriers which in its turn depends on temperature.

 \section{The energy resolution and threshold of GDA.}

  The energy resolution of semiconductor detector  will be given by
\beq
\Delta E=\sqrt {(\Delta E_{int})^2+(\Delta E_{el})^2}
\eeq

  where $\Delta E_{int}$ is the intrinsic energy resolution of detector which is
 defined by statistical fluctuation in the number of charge carriers created
in detector sensitive volume, $\Delta E_{el}$ is the energy resolution which
 is defined by associated electronics. In the case of GDA these two terms
 will be of form
\beq
\Delta E_{int}=2.34\sqrt{\varepsilon E(F+f)K^2}
\eeq

 and
\beq
\Delta E_{el}=\frac{4.52\varepsilon}{e}\sqrt{\frac{0.6kT}{\tau S} C^2+
kT\tau\left [\frac{1}{R_{\Sigma}}+\frac{e}{2kT}(I_s+I_bfK^2)\right]}
\eeq
 where $\varepsilon$ is the energy to create one pair of charge carriers, $E$
 is the energy deposited in the detector, $F$ is the Fano factor, $f$ is the
excess noise factor due to the fluctuation of multiplication, $K$ is the
amplification factor, $e$ is the electron charge, $T$ is the absolute
temperature of the resistors, $C$ is the total capacitance presented at the
 input of preamplifier, $\tau$ is the time constant of the RC circuits of the
preamplifier, $S$ is the steepness of the field effect transistor
characteristic, $R_{\Sigma}$ is the resistance at the input of the
preamplifier, $I_s$ is the detector leakage surface current and $I_b$ is the
 detector leakage balk current due to the thermal generation of charge
 carriers. According to the calculation for GDA with K$>$10 one must take
into account in the formula for  $\Delta E_{el}$  only the last term due to
the bulk leakage current and the formula (7) for GDA  can be rewritten as
\beq
\Delta E\approx 2.36\cdot K \sqrt {\varepsilon E(F+f)+10^4I_b\tau
f} \eeq where $\varepsilon$ and $E$ is in eV , $I_b$ is in nA and $\tau$ is
in $\mu sec$.  The GDA energy threshold is defined by $I_{b}$ or more exactly
 by the last term in (10):  \beq E_{th} \ge 2.36 \sqrt{ 10^4\cdot I_b\tau
f} \eeq

One can estimate the $E_{th}$ for microstrip planar detector from HPGe of
 volume 100 $cm^3$ with internal amplification: at N=$10^{10}$$cm^{-3}$,
$I_b$=0.01 nA per one strip, $\tau$=0.5 $\mu sec$ and $f$=0.5 : $E_{th}\ge$ 12
eV.

  The dependance of
   relative energy resolution  $\Delta$E/E on energy in the more interesting
 energy range $50\div5000$ $eV$ for GDA is shown in table.  \begin{table} [h]
\caption{ The dependance of relative energy resolution $\Delta$E/E on
energy } \begin{center} \begin{tabular}{|c|c|c|c|c|c|c|c|} \hline $E(eV)$ & 50
& 200 & 400 & 600 & 800 & 1000 & 5000\\ \hline $\Delta E/E(\%)$ & 57 & 25 &
17 & 13 & 11 & 10 & 4.3\\ \hline \end{tabular} \end{center} \end{table}\\ The
 internal amplification of GDA degrades the performance somewhat, but such
 energy resolution of GDA is adequate to the investigation of above problems.
  It is interesting to note that common planar detector from HPGe of volume
100 $cm^3$ produced by "Canberra" (type GL3825R) has relative energy
resolution  about 8$\%$ at E=5900 eV.

\section{  The prospects of GDA use.}

 Presently germanium detectors are in considerable use in low background
measurements for high purity of germanium crystals: their content of
radioactive impurity does not exceed $10^{-13} g/g$ ${\cite{2beta}}$. If the
internal amplification is realized in germanium detectors and their threshold
is lowered to several  $eV$  the possibility for their use will
considerably increase.
Firstly, the lowering of the detector threshold
 makes it possible to extend the kinematical domain of investigations,
 secondly, it enables in some cases significantly to decrease the level
 of background. Following are brief discussions  of prospects of GDA using
 for neutrino magnetic moment ( $\mu_{\nu}$ ) measurement, for investigation of
neutrino coherent scattering by nuclei and for search for "light" WIMPs.

  {\bf Neutrino magnetic moment(NMM) measurement.}

    A laboratory measurement of the NMM is based on its contribution
to the (anti)neutrino - electron scattering. For a non-zero NMM the
differential over the kinetic energy ${ T}$ of the recoil electron
cross section ${ d \sigma/dT}$ is given by the sum of the weak
interaction cross section and the electromagnetic one.
At a small recoil energy the weak part practically constant, while
the electromagnetic one grows as ${1/T}$ towards low energies.
For improving
the sensitivity to $\mu_{\nu}$ it is necessary to lower the threshold
for detecting the recoil electrons as far as the background
conditions allow.
Now there is
 number of the projects dedicated to the NMM measurements with detectors of
mass $2\div1000$ kg and with thresholds of $3\div500$ KeV. Authors of
proposal ${\cite {star}}$
 are going to use germanium detector with mass of 2 kg and threshold of 3 KeV
 to achieve for two years reactor measurement the limit on $\mu_{\nu}\sim
3\times 10^{-11} \mu_B$, with $\mu_{B}=e/(2m_{e})$ being the Bhorh magneton.
 The use of GDA with internal
 amplification in this experiment would provide possibility to achieve limit
on $\mu_{\nu}$ about 2$\times 10^{-11}\mu_B$.  However, one can not
significantly low the GDA threshold in the reactor experiment for two
reasons. Firstly, one must take into account atomic binding effect for
electrons. The total $\tilde\nu_e$ cross section will be 30$\%$ less due to
the binding effects ( in the energy range $200\div3000$ eV ). The secondly at
 T$<$500 eV the effect neutrino coherent scattering (NCS) by germanium nuclei
 will surpass the effect of $\tilde\nu e$ scatttering and NCS will present
 physical background relative to the $\tilde\nu e$ scattering. Nevertheless
 the low GDA threshold can be used in full measure if one uses artifical
 neutrino source (ANS) instead of reactor.  Nowadays there are several
 proposals on development ANS with activity $5\div40$ MCi ${\cite{kar}}$.
 Although the development and construction of the ANS is rather expensive and
 challenging problem the use of ANS offers some advantages over the reactor
 based experiments:

 1) ANS can provide significantly higher neutrino flux density (up to
  $10^{15}$ 1/$cm^2\cdot s$),

 2) ANS can be used in deep underground laboratory , where level of background
    is significanly lower than in shallow box near the reactor,

 3) one can use the optimum ratio of effect and background measurement times,
    that is impossible in the reactor experiment,

 4) the uncertainities of the neutrino spectrum and flux density are small.

 Presently the more appropriate ANS for NMM measurement is a tritium source
 proposed in ${\cite{trof}}$ for measurement of NMM by means of low threshold
cryogenic silicon detector. The end-point energy of tritium beta spectra is
 18,6 KeV and for maximum recoil electron energy a kinematical limitation
 gives:  \beq T_{max} = \frac{2E_\nu^2}{2E_\nu+ m_e c^2} = 1260 eV. \eeq If
 one uses tritium ANS with activity 40 MCi then the neutrino density flux
 will be 6$\times 10^{14}$ 1/$cm^2\cdot s$ in the detector position. The
 limit on NMM about 3$\times 10^{-12}\mu_B$ can be achieved during one year
 measurement time with silicon detector mass of 3 kg in the energy range
 $1\div300$ eV. The very same limit on NMM can be achieved by use the GDA with
 mass of 3 kg during one year of measurement time in the energy range
 $30\div500$
 eV. In the GDA case the count rate due to electromagnetic neutrino
 interaction will be 0.24 event/d (at $\mu_{\nu}=3\times 10^{-12} \mu_B$),
 the count rate due to weak interaction will be one order of magnitude less.
 Now let us consider the background count rate. If one takes into account that
 $\sigma_{EM}\sim ln \:({T_{max}}/{T_{min})}$ then the value of $\sigma_{EM}$
 in the energy range $3\div50$ KeV($\Delta E_1$) and $30\div500$ eV($\Delta
 E_2$) will be the very same.  The best level of background for germanium
 detector was achieved in ${\cite{klap}}$ and it is equal 0.1
 event/keV$\cdot$kg$\cdot$d.  If one assume that level of background does not
 depend on energy in the energy range of interest then the level of
  background in the second energy interval will be two order of magnitude
 less since $({\Delta E_2}/{\Delta E_1})\sim 10^{-2}$ and will be 0.15
 event/d.  In addition in the case of tritium ANS there is no physical
 background due to the NCS since in this case the maximum recoil energy of
 germanium nuclei $T_{max}^n$ is equal 0.01 eV.

{\bf Neutrino coherent scattering by nuclei.}

 The use of GDA can open up the possibility to investigate the coherent
 neutrino scattering  by nuclei. This interaction is of great importance
 into interstellar processes  and up to now did not observed in laboratory
 because the very low kinetic energy transferred to nucleus in the process of
 neutrino-nucleus scattering. For germanium nucleus and reactor antineutrino
 spectrum the maximum kinetic recoil energy is about $T_{max}^n$=2500 eV and
 taking into account  quenching effect only one third of this energy are
 imparted on ionization ${\cite {bern}}$.
 So one needs germanium detector with threshold significantly lower 800 eV
 for investigation of NCS by germanium nuclei. Such low threshold can be
 provided by GDA having internal amplification about 10$^2$. The differential
 cross  section of NCS by nuclei ${\cite{blas,dodd}}$ can be expressed by :
 \beq
 \frac{d\sigma_W^c}{dT}\simeq\frac{G_F^2}{4\pi}N^2M\left(1-
 \frac{MT^n}{2E_{\nu}^2}\right)
 \eeq
 \beq
 \frac{d\sigma_{EM}^c}{dT}\simeq\left(\frac{\mu_{\nu}}{\mu_B}\right)^2
 \frac{\pi\alpha^2}{m_e^2}Z^2\left(\frac{1}{T^n}\right)
 \eeq
 where M,N and Z are the mass, neutron and charge numbers of nuclei
 correspondingly and T$^n$ - nucleus recoil energy. If one uses GDA with
 mass of 3 kg and threshold 30 eV in the reactor experiment the count
 rate due NCS in the energy range $30\div300$ eV will be near 100 event/d
at antineutrino intensity
 $2\times 10^{13}$ $\nu/cm^{2}\cdot s$ due to the weak interaction.
The count rate due to the electromagnetic interaction (at $\mu_{\nu} = 2
\times 10^{-11} \mu_{B}$) will be 0.45 event/d. The
level of detector background will be only 0.1 event/d if one uses the above
estimations of detector background, taking into account that the energy range
is equal 0.3 KeV in this case.  So one can see that use of GDA for
measurement NCS makes it possible :  \begin{itemize} \item{ to improve the
weak interaction constants,} \item{ to investigate the neutrino oscillations
by alternative way,} \item{ to make more precise the quenching factor for
germanium nuclei at low energy transfer which is of interest for dark matter
 experiments.} \end{itemize}

 {\bf The GDA use in the Dark Matter Experiment.}

 Presently in Dark Matter Experiments the main efforts are directed at
 decreasing of the background and lowering of the energy threshold of the
 detector, since for more appropriate dark matter candidate- WIMPs or weakly
 interacting massive particles - the expected count rate for WIMPs-nuclear
 scattering is in the range $ 0.001\div1.0$ event/kg$\cdot$d and expected
recoil energy $T^n$ lies in the wide energy range beginning of some tens $eV$
 and above.  Significant success was achieved by CDMS (Cryogenic Dark Matter
 Search) collaboration in radiation background decreasing ${\cite{cdms}}$.
The rejection of 99$\%$ of photon background was achieved use detectors which
 simultaneously measure the recoil energy in both- photon and charge mediated
 signals. However, this method is effective at detector threshold above
 15 KeV.
 So experiments for search for WIMPs with mass lower than 10 Gev/$c^2$
 (at very low T) are carried out now by CRESST (Cryogenic Rare Event Search
 with Superconducting Thermometers) collaboration ${\cite{cres}}$ which plans
to use cryogenic detector including four 250 g sapphire detectors with
threshold 500 eV. The using of several GDA modules of mass 1 kg each
 with threshold about 30 eV would be very effective in this investigations.

   The investigation of Solar Neutrino Problem with use GDA can provide
 possibility to detect simultaneously the whole neutrino spectrum ${\cite{
 blas}}$ but one
 needs of course to use large mass detector in this case.
 It is beleived that GDA being developed will find also use in the applied
 fields.

   The authors would like to thank V.S.Kaftanov for his participation in
 this work and Yu.Kamishkov for his interest to this work and for
 enlightening discussions.

\newpage
\thispagestyle{empty}
\begin{figure}
\epsfysize=6.0in
\epsfbox[20 0 520 500]{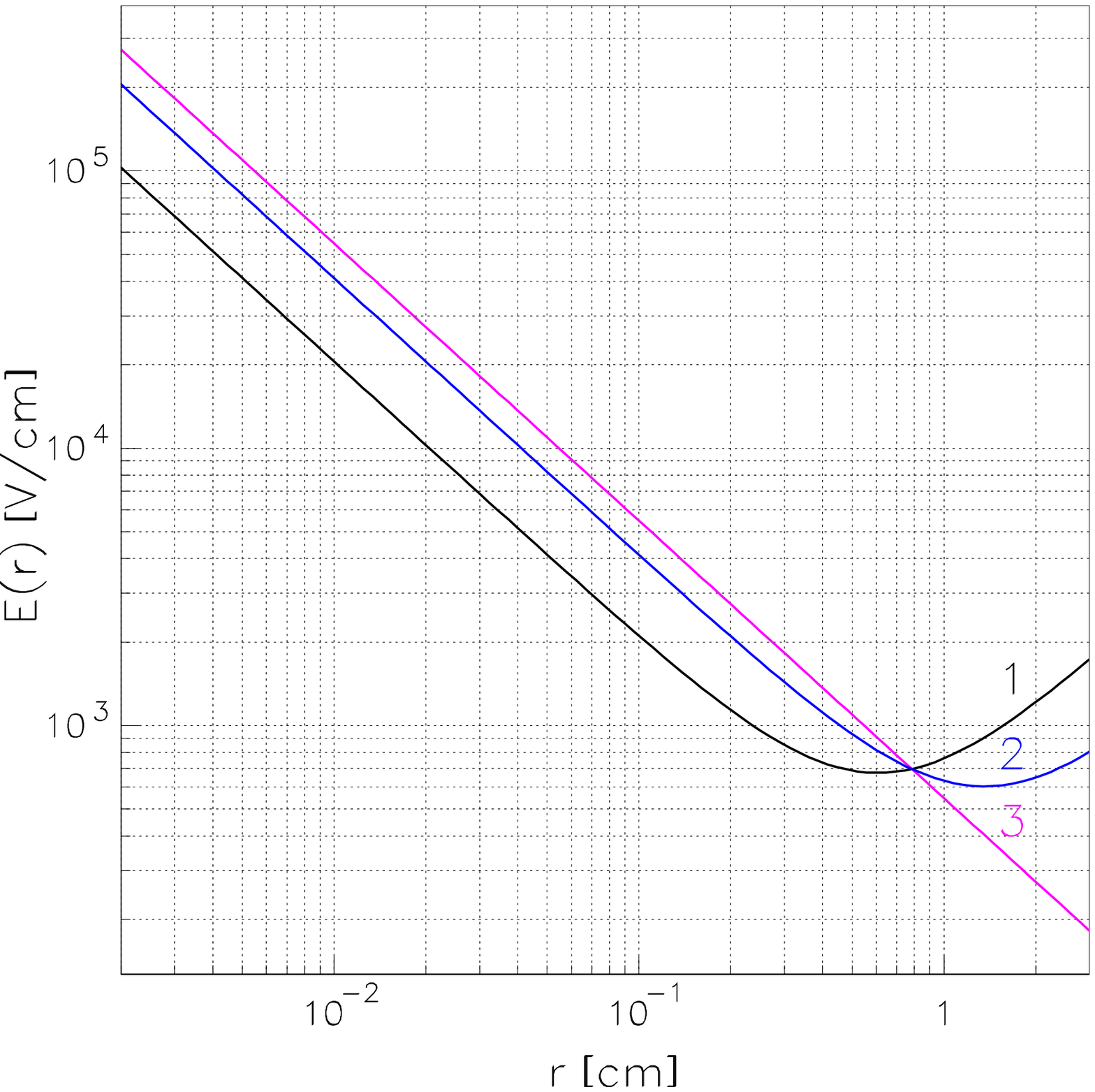}
\caption{
Dependence of the electric field on $r$ for axial  detector from HPGe
n-type at V = 4000 $V$, $R_{1}~=~0.002~cm$, $R_{2}~=~3.0~cm$ and at
 different impurity concentrations: {\bf 1} - $N=10^{10}~ cm^{-3}$,
     {\bf 2} - $N~=~4\times10^{9}~ cm^{-3}$,  {\bf 3} - $ N~=~0$ ( volume
charge is absent).
 } \end{figure} \newpage \thispagestyle{empty}
\begin{figure}
\epsfysize=6.0in
\epsfbox[20 0 520 500]{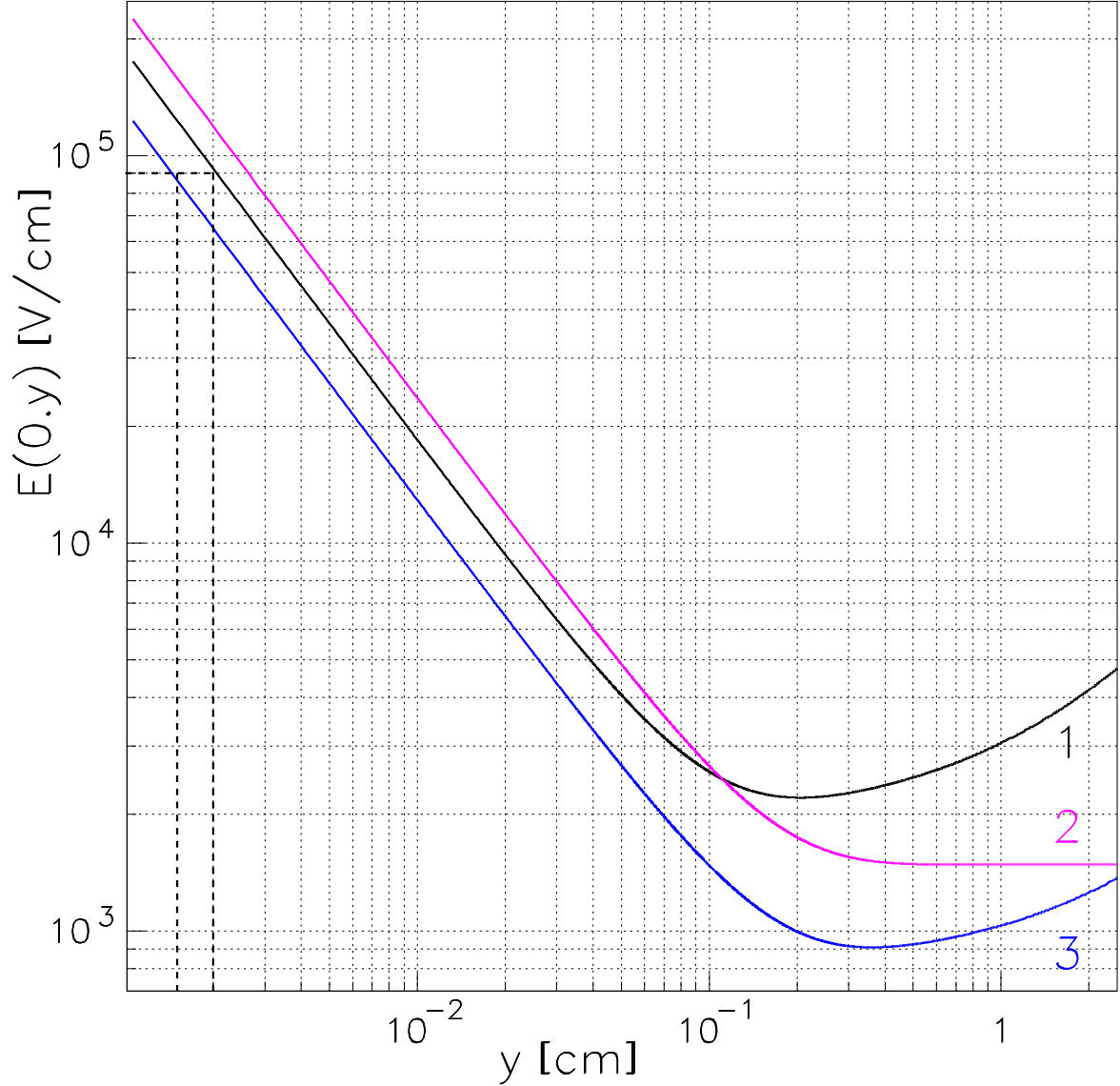}
\caption{
Dependence $E(0,y)$ on $y$ for planar detector from  HPGe n-type with
d~=~20~ microns at V = 4000 $V$ and at different values of the others
parameters:  {\bf 1} - $N=10^{10}~ cm^{-3}$, $L~=~1.5~ cm$ and $s=0.3~cm$;
 {\bf 2} - $N=0$ ( volume charge is absent ), $L=2.0~cm$ and $s=0.5~cm$: {\bf
3} - $N=2\times 10^{9}~ cm^{-3}$, $L=3.0~cm$ and $s=0.5~cm$.  For ({\bf 1})
  and ({\bf 3}) the length of avalanche region is equal 10 and 5 microns,
accordingly.  } \end{figure} \newpage \thispagestyle{empty} \begin{figure}
\epsfysize=8.0in \epsfbox[20 40 526 727]{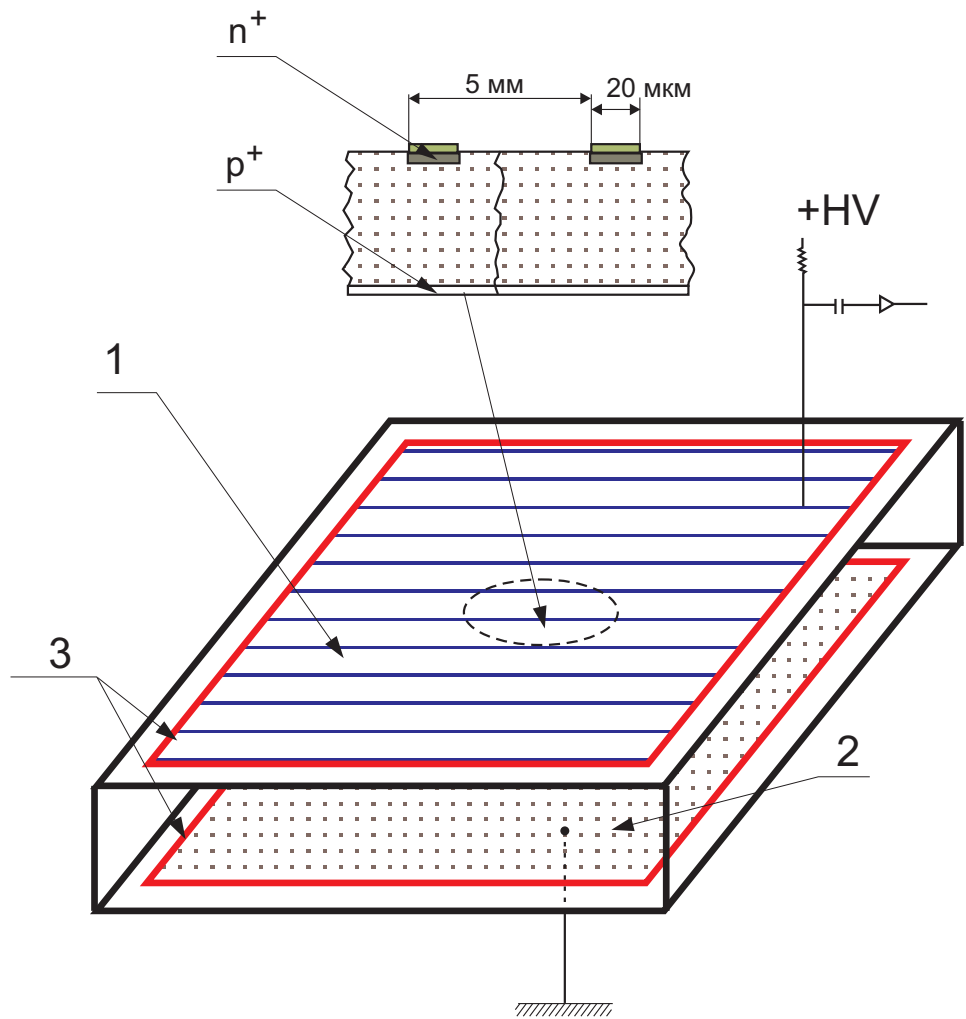} \caption{ Germanium
detector with internal amplification.  1 - anode strips, 2 - cathode, 3 -
guard electrodes, the scheme  of $n^{+}$  and $p^{+}$ - layers are shown in
the upper part.}
\end{figure}
\end{document}